\documentclass[a4paper,oneside,fleqn,twocolumn,showkeys,showpacs]{revtex4}

\usepackage{graphicx}
\usepackage{amsmath}
\newcommand{\R}{\text{${\rm{I\!R}}$}}

\begin{document}
\title{A set of basis functions to improve numerical calculation of Mie scattering in the Chandrasekhar-Sekera representation}

\author{\firstname{Alexandre} Souto \surname{Martinez}}
\email{asmartinez@ffclrp.usp.br}
\affiliation{Faculdade de Filosofia, Ci\^encias e Letras de Ribeir\~ao Preto, \\
             Universidade de S\~ao Paulo \\
             Av. Bandeirantes, 3900 \\
             14040-901, Ribeir\~ao Preto, SP, Brazil. }
\affiliation{National Institute of Science and Technology in Complex Systems (LNCT-SC)}             

\author{\firstname{Tiago} Jos\'e \surname{Arruda}}
\email{tiagoarruda@pg.ffclrp.usp.br}
\affiliation{Faculdade de Filosofia, Ci\^encias e Letras de Ribeir\~ao Preto, \\
             Universidade de S\~ao Paulo \\
             Av. Bandeirantes, 3900 \\
             14040-901, Ribeir\~ao Preto, SP, Brazil. }
\date{\today}

\begin{abstract}
Numerical calculations of light propagation in random media demand the multiply scattered Stokes intensities to be written in a common fixed reference.
A particularly useful way to perform automatically these basis transformations is to write the scattered intensities in the Chandrasekhar-Sekera representation.
This representation produces side effects so that numerical tests are necessary to deal with the limiting situations of the small-particle (Rayleigh) and forward/backward  scattering.
Here a new set of basis functions is presented to describe the scattering of light by spherical particles (Mie scattering) in the Chandrasekhar-Sekera representation.
These basis functions can be implemented in a new algorithm to calculate the Mie scattering amplitudes, which leads straightforwardly to all the scattering quantities.
In contrast to the traditional implementation, this set of basis functions implies to natural numerical convergence to the above mentioned limiting cases, which are thoroughly discussed.
\end{abstract}

\keywords{Mie scattering, Chandrasekhar-Sekera representation, multiple scattering, random media, radiative transfer equation, Monte Carlo simulation, computation technics}
\pacs{03.65.Nk, 42.25.Dd, 03.50.De,02.70.-c}


\maketitle

\section{Introduction}


The scattering of an electromagnetic plane wave by a spherical (homogeneous, isotropic and optically linear material) particle  of arbitrary size is known as Mie scattering~\cite{vandehulst_book,kerker_book,bohren_book,mishchenko_book}.
The numerical calculation of this scattering is relevant to several fields such as remote sensing (meteorological optics, radar detection of raindrops, lidar detection of clouds etc.)~\cite{mishchenko_remote_book}, optical particle characterization (inverse problems)\cite{wolf,klett,chahine,li}, photonic band gaps materials (PBGs) \cite{sigalas,zhang} etc.
More recently, magnetic Mie scattering has attracted the attention of researchers~\cite{kerker:1983,pinheiro_prl1_2000,pinheiro_prl2_2000,pinheiro_j3m_2001,pinheiro_bjp_2001,mehta_2006,pradhan_2006,patel_2006,tiago-sphere,tiago-cylinder}.


In optical dense media, the light scattered by a particle is successively rescattered and the electromagnetic wave phase coherence may not be totally destroyed by the particle configuration averages producing interesting effects due to phase correlation~\cite{sheng_book}.
The role of numerical multiple scattering description is twofold: it can be viewed as a tool in the comprehension of more fundamental aspects (memory of incident polarization state~\cite{zhu:89,martinez:3:94,alfano_2005,xu_alfano_2005,gorodnichev_2006}, weak localization~\cite{lagendijk_1996,poan_1998,mishchenko_multiple_book}) or as a tool for random media characterization (light scattering by biological tissues~\cite{ghosh_2002,zimnyakov_2002,sankaran_2002,sun_2003,ghosh_2003,gupta_2003,alfano_2004,liu_2005,angelsky_2005,gupta_2005,itoh_2005}, for instance).
These procedures demand considerable numerical efforts and it is desirable to have very efficient and robust (covering a wide range of values in the parameter space) codes to  perform such tasks.


To deal with numerical multiple scattering simulation, basis transformations must be performed to represent the local scattered Stokes intensities into a common fixed basis, the laboratory basis.
For instance, this procedure is successively repeated in multiple scattering of light in a Monte Carlo scheme.
Other than being cumbersome, these basis transformations may increase the propagation of round-off errors in the numerical simulations and may ask for complementary tests in limiting situations.


These basis transformations may be implicitly considered writing all the wave\-vectors in the same laboratory basis.
This is the Chandrasekhar-Sekera representation~\cite{chandrasekhar_book,sekera}, which has been employed either in a radiative transfer equation calculation~\cite{cheung} as well as in a Monte Carlo scheme~\cite{martinez:5:1994}.
The drawback of this representation (not stressed in the literature) is the introduction of new difficulties to numerical calculation, notably the non-commutation of the two limiting cases: \emph{(i)} small particle size compared to the wavelength (important to polydispersion calculations), and \emph{(ii)} forward/backward scattering events (important to variance reduction in a Monte Carlo schemes and radar/lidar detection).



In this paper the problem of limits non-commutability is pointed out and a new set of basis functions for Mie scattering calculation is proposed to the implementation of an algorithm.
This algorithm naturally includes the small particle size and backward/forward scattering.
The presentation is divided as follows.
In Section~\ref{sec:Scattering_plane_representation}, a brief review of Mie scattering in the scattering plane representation is presented to set up the notation.
Also, an original expansion of the scattering amplitudes up to fourth-order on the cosine of the scattering angle is calculated.
In Section~\ref{sec:Chandrasekhar-Sekera_representation}, the Chandrasekhar-Sekera representation is reviewed.
It is pointed out in Section~\ref{sec:limit_problems} the non-commutability of the limiting cases.
In Section~\ref{sec:the_new_set_of_basis_functions}, a new basis functions are presented to calculate Mie scattering in the Chandrasekhar-Sekera representation and its implementation for numerical calculation is discussed.
Concluding remarks are presented in Section~\ref{sec:conclusion}.

\section{Mie Scattering}
\label{sec:Scattering_plane_representation}


Let us consider a non-absorbing medium with (real) refractive index $n_m$ ($n_m\in \R$), a sphere of radius $a$ and (complex) refractive index $n_s$ ($n_s \in \mathcal{C}$) to take emission or absorption into account.
Here we use the same scattering approach as van de Hulst~\cite{vandehulst_book} and Kerker~\cite{kerker_book}, so that ${\rm Im}(n_s)\leq0$.
The origin of the laboratory frame $(\hat{x}, \hat{y}, \hat{z})$ is placed at the center of a spherical particle.
An incident monochromatic (plane) wave, with electric field $\vec{E}_0$ and wavelength $\lambda$ (consequently the wavenumber $k = 2 \pi / \lambda$), propagates along the $z$-direction with wavevector $\vec{k}_{0} = k \hat{z}$ and it is  scattered by the sphere.
The interest is on the spherical scattered field $\vec{E}_1$ along the $\vec{k}$ direction, defined by the spherical angles $\theta$ and $\phi$ according to the laboratory frame.
The scattering plane is formed by the vectors $\vec{k}_0$ and $\vec{k}$.
Notice that the scattering plane cannot be univocally defined for backward/forward scattering cases.
Although the magnetic field (of the electromagnetic wave) does not explicitly appear in the calculation, it has not been neglected since it can always be obtained from the electric fields (due to the consideration of transverse wave).


\subsection{Scattering Plane Representation}


In a distance $R$ from the center of the sphere, with $(ka)^2/kR \ll 1$ (far-field approximation),  the scattered field is nearly transverse.
The electric field lies on the plane orthogonal to $\hat{k}$ formed by the orthogonal directions $\hat{\theta}$ and $\hat{\phi}$, which  form the spherical basis.
Using $\imath = \sqrt{-1}$ for the imaginary part, the scattered field in the spherical basis is written as:
\label{eq:sct_plane}
\begin{equation}
\label{eq:ef_sct_plane}
\left[
\begin{array}{c}
E_\theta \\
E_\phi
\end{array}
\right]_1 = - \imath \; \frac{e^{- \imath k R}}{k R} \;
J(\mu, \phi) \;
\left[
\begin{array}{c}
E_x \\
E_y
\end{array}
\right]_0 \; ,
\end{equation}
with $\mu = \cos \theta$, where $\theta$ is the scattering angle and $\phi$ is the azimuthal angle.
The Jones matrix is the product of a diagonal  matrix (because of the scatterer spherical symmetry), known as scattering matrix with a rotation matrix:
\begin{equation}
\label{eq:jones_spr}
J(\mu, \phi)  =
\left[
\begin{array}{cc}
S_{\parallel}(\mu) & 0                      \\
0                          & S_{\perp}(\mu)
\end{array}
\right] \;
\left[
\begin{array}{cc}
  \cos \phi & \sin \phi         \\
- \sin \phi & \cos \phi
\end{array}
\right] \; .
\end{equation}
The rotation matrix projects the incident electric field (given in the laboratory frame) to the parallel and perpendicular directions relative to the scattering plane.
The scattering matrix then alters the field values via $S_{\parallel}(\mu)$ and $S_{\perp}(\mu)$, which are the parallel and perpendicular scattering amplitudes (relative to the scattering plane), respectively,
\begin{eqnarray}
S_{\parallel}(\mu) & = & \sum_{n = 1}^{\infty} \frac{2n + 1}{n(n+1)} \; [a_n \tau_n(\mu) + b_n \pi_n(\mu)] \; ,
\label{eq:s1s2:1} \\
S_{\perp}(\mu)     & = & \sum_{n = 1}^{\infty} \frac{2n + 1}{n(n+1)} \; [a_n \pi_n(\mu) + b_n \tau_n(\mu)] \; ,
\label{eq:s1s2:2}
\end{eqnarray}
where $a_n$ and $b_n$ are the Mie coefficients~\cite{vandehulst_book,kerker_book,kerker:1983}, and
\begin{eqnarray}
\pi_n(\mu) & = & \frac{P_n^{(1)}(\mu)}{\sqrt{1 - \mu^2}} \; ,
\label{eq:pi} \\
\tau_{n}(\mu)  & = & \frac{d P_n^{(1)}(\mu)}{d \theta} =  \mu \pi_n(\mu) - (1 - \mu^2) \pi^{\prime}_n(\mu) \; ,
\label{eq:tau}
\end{eqnarray}
with $\pi^{\prime}_n(\mu) = {\rm d} \pi_n(\mu) / {\rm d} \mu$ and  $P_n^{(1)}(\mu)$ being the $n^{\mbox{th}}$ Legendre polynomial of first order.
In practice the summations in $n$ in Eqs. (\ref{eq:s1s2:1}) and (\ref{eq:s1s2:2}) must be performed to $n_{\max} = ka + 4({ka})^{1/3} + 2$~\cite{nussenzveig_book,wiscombe:1980}.

The Mie scattering is dependent only on three quantities: the size parameter $ka$, (complex) relative refractive index $m = n_s/n_m$ ($m \in \mathcal{C}$) and $\widetilde{m} = m/\widetilde{\mu}$ (relative impedance), where $\widetilde{\mu} = \mu_s/\mu_m$ is the relative (sphere/medium) complex magnetic permeability~\cite{kerker:1983,pinheiro_prl1_2000,pinheiro_prl2_2000,pinheiro_j3m_2001,pinheiro_bjp_2001,mehta_2006,tiago-sphere}.
Notice that sphere and medium refractive indices are $n_s = \sqrt{\epsilon_s \mu_s/(\epsilon_0 \mu_0)}$ and $n_m = \sqrt{\epsilon_m \mu_m/(\epsilon_0 \mu_0)}$, where $\epsilon_0$ and $\mu_0$ are the vacuum permitivity and permeability, respectively.
The numerical calculation of this scattering event consists of two parts: the first one involving wavelength, relative refractive indices and size of the scatterer (Mie coefficients $a_n$ and $b_n$, which directly leads to several cross-sections) and the second one involving the scattering angular dependence (scattering amplitudes $S_{\parallel}$ and $S_{\perp}$, which leads to the phase function).


The calculation of Mie coefficients $a_n$ and $b_n$ depends on $ka$, $m$ and $\widetilde{m}$ through the calculation of spherical Bessel, Neumann, and Hankel functions and their first derivatives with respect to their argument (either $ka$ or $mka$).
The difficulty in this calculation is that the recurrence relationships for complex arguments of the Bessel functions are not stable.
This problem is solved writing the spherical functions in the ratio form~\cite{grehan} for Mie coefficients and with the use of the continued fraction method developed by Lentz~\cite{lentz}.

If a desired precision in the results is known in advance, it is possible to use the recurrences for $a_n$ and $b_n$ developed by Bohren~\cite{bohren:2} and implement the scheme proposed by Cachorro~\cite{cachorro}.
Further improvements to this calculation have been compiled and studied in detail in Ref.~\cite{du_2004}.


The second part of the calculation concerns the scattered field.
It depends on the relative position of the detector with respect to the scatterer and source and also on the distance between scatterer and detector.
The scattering amplitudes, which depend on the azimuthal angles, cosine of the scattering angle $\mu$ and on the Mie coefficients, can be efficiently calculated using the algorithm created by Wiscombe~\cite{wiscombe:1980}.

Observe that the total scattering cross section is obtained by means of the optical theorem, which is obtained expanding the scattering amplitudes up to second order on the scattering angle.

\subsection{Scattering Amplitudes}
\label{sec:Expansion_scattering_amplitudes}

The expansions of the scattering amplitudes are presented up to the fourth-order on the scattering angle.
These expansions have been explicitly included since they are not easily found in the current literature.
As shown in Ref.~\cite[p. 73]{kerker_book}, the Legendre polynomials can be written as: $P_n(\cos \theta)  =  \sum_{p = \delta, \delta + 2}^{n} \mu_n^p \cos^p \theta$, where $\delta = 0(1)$ if $n$ is even (odd) and $\mu_n^p   =  (-1)^{(n - p)/2} 1 \cdot 3 \cdot 5 \cdots (n + p - 1)/\{2^{(n - p)/2}[(n - p)/2]! p !\}$
so that: $P_n^{(1)}(\cos \theta) = \sin \theta  {\rm d} P_n (\cos \theta)/{\rm d} \cos \theta$ and Eqs.~(\ref{eq:pi})~and~(\ref{eq:tau}) are rewritten as:
$\pi_n (\cos \theta)  =  \sum_{p = \delta, \delta + 2}^{n} \mu_n^p p \cos^{p-1} \theta$, $\tau_n(\cos \theta)  =  \cos \theta \pi_n(\cos \theta) - \sin^2 \theta \pi_n^{\prime}(\cos \theta)  =  \sum_{p = \delta, \delta + 2}^{n} \mu_n^p p \cos^{p-2} \theta (1 - p \sin^2 \theta)$.

\subsubsection{Expansion along the Forward/Backward Direction}

Using the following sum rules:
\begin{eqnarray}
H^{(\pm)}_n   =  \sum_{p = \delta, \delta + 2}^{n} (\pm 1)^{p-1} \mu_n^p p
                       & = & \frac{(\pm 1)^{n+1} n(n + 1)}{2} \; ,
\label{eq:Hn} \\
\nonumber
K_n^{(\pm)} =   \sum_{p = \delta, \delta + 2}^{n} (\pm 1)^{p-1} \mu_n^p p^2
                        & = & \frac{H_n^{(\pm)}}{2}[1 + (\pm 1)^{n+1} H_n^{(\pm)}] \\
                        & = & \frac{\left[n(n+1) + 2\right]}{4} \; H_n^{(\pm)}
\label{eq:Kn} \; ,
\end{eqnarray}
and expanding the $\pi_n(\mu)$ and  $\tau_n(\mu)$ functions up to $O(\theta^4)$ around the back and forward scattering directions
\begin{eqnarray}
\frac{\pi_n (\mu \rightarrow \pm 1)}{H^{(\pm)}_n} & = &  1 + \left[1 -   \frac{n(n+1)}{2} \right] \; \frac{\theta^2}{4} + \ldots \; , \\
\frac{\tau_n(\mu \rightarrow \pm 1)}{\pm H^{(\pm)}_n} & = &  1 +  \left[1 -   \frac{3n(n+1)}{2} \right] \; \frac{\theta^2}{4}  + \ldots  \; ,
\end{eqnarray}
one obtains the forwards/backwards scattering amplitudes:
\begin{eqnarray}
S_{\parallel}(\mu \rightarrow \pm 1) & = & S(\pm 1) + L^{(\pm)} \theta^2   + \ldots \; ,
\label{eq:scat_amp:1} \\
S_{\perp}(\mu \rightarrow \pm 1) & = &    \pm S(\pm 1) + P^{(\pm)} \theta^2   + \ldots \; ,
\label{eq:scat_amp:2}
\end{eqnarray}
where
\begin{eqnarray}
\label{eq:s0}
S(\pm 1) & = & \sum_{n = 1}^{\infty} S^{(\pm)}_n (b_n \pm a_n) \; , \\
8 L^{(\pm)} & = & \sum_{n = 1}^{\infty} S^{(\pm)}_n [ 2( b_n \pm  a_n) -  n(n+1) (b_n \pm 3 a_n) ] \; , \\
8 P^{(\pm)} & = & \sum_{n = 1}^{\infty} S^{(\pm)}_n [ 2( a_n \pm  b_n) -  n(n+1)  (a_n \pm 3 b_n)]  \\
S^{(\pm)}_n & = &  \frac{2n + 1}{n(n+1)} H^{(\pm)}_n \; .
\end{eqnarray}

From Eqs.~(\ref{eq:scat_amp:1}) and~(\ref{eq:scat_amp:2}), one obtains the following properties for the exact forward amplitude, which is used to calculate the total scattering cross section using the optical theorem: $S_{\perp}( 1)  =   S_{\parallel}(1)$ and backward scattering  amplitude, which gives the backscattering (radar) cross section~\cite{vandehulst_book,kerker_book,bohren_book,mishchenko_book}, $S_{\perp}(-1)  =  -S_{\parallel}(-1)$.
For future use, let us define the following quantity
\begin{eqnarray}
\nonumber
\Delta^{(\pm)} & = & \pm P^{(\pm)} - L^{(\pm)} \\
                          & = & \frac{1}{4}  \sum_{n = 1}^{\infty} (2n + 1) H^{(\pm)}_n   (\pm a_n - b_n) \; ,
\label{eq:delta}
\end{eqnarray}
where $H^{(\pm)}_n$ is given by Eq.~(\ref{eq:Hn}).

\subsubsection{Expansion for Small Particle Size}

The general Mie coefficients for the magnetic case has been pointed out in Ref.~\cite{kerker:1983} and
in the limit of small-particle, the size parameter $ka \ll 1$, they are~\cite{pinheiro_prl1_2000}:
\begin{eqnarray}
a_1 & = & \frac{\gamma}{2} \; \frac{ka - 2 \widetilde{m} A_1(mka)}{ka +  \widetilde{m} A_1(mka)} \\
b_1 & = & \frac{\gamma}{2} \; \frac{\widetilde{m} ka - 2  A_1(mka)}{\widetilde{m} ka +  A_1(mka)} \; ,
\end{eqnarray}
where $\gamma  =  -2 \imath (ka)^3/3$, $A_1(z)  = \psi_1(z)/\psi_1'(z)$  and $\psi_1(z) = z j_1(z)$ is the spherical Bessel-Riccati function ($j_1(z) = \sin z /z^2 - \cos z/z$) and prime means the derivative with respect to the argument of the function.
This is the general magnetic scattering by small particles.

To ensure the Rayleigh scattering, the wavelength must also be small inside the scatterer  $|m|ka \ll 1$ leading to~\cite{kerker:1983}
\begin{eqnarray}
a_1 & = &  \gamma \; \frac{1 - \widetilde{\epsilon}}{2 + \widetilde{\epsilon}} \\
b_1 & = &  \gamma \; \frac{1 - \widetilde{\mu}}{2 + \widetilde{\mu}} \; ,
\end{eqnarray}
where $\widetilde{\epsilon}$ and $\widetilde{\mu}$ are relative (sphere to medium) permittivity and permeability.
Notice that the whole magnetic dependence is in the $b_1$ term, which vanishes for non-magnetic scattering ($\widetilde{\mu} = 1$).

Next the auxiliary quantities are written in the small-particle limit
\begin{eqnarray}
S(\pm 1) & = & \frac{3}{2} (b_1 \pm a_1) \; , \\
L^{(\pm)} & = & \frac{\mp 3}{4} \; a_1  \; , \\
P^{(\pm)} & = & \frac{\mp 3}{4} \; b_1 \ , \\
\Delta^{(\pm)} & = & \frac{3}{4} ( \pm a_1 - b_1) \; ,
\end{eqnarray}
where near forward/backward scattering, the parallel scattering amplitude is influenced only by an electric component $L^{(\pm)}$ and the perpendicular one only by the magnetic correction $P^{(\pm)}$.

\section{Chandrasekhar-Sekera Representation}
\label{sec:Chandrasekhar-Sekera_representation}

\begin{figure}
\begin{center}
\includegraphics[width=\columnwidth]{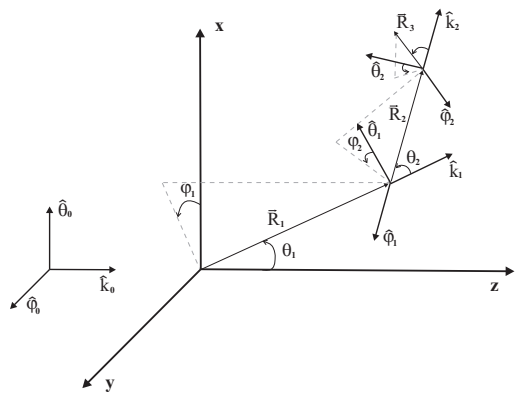}
\end{center}
\vspace{-7cm}
\caption{Sequence of three scattering events where the Chandrasekhar-Sekera basis are explicitly shown.}
\label{fig:1}
\end{figure}

With the scattering plane representation one can express directly the photon ``history'' from the source to the detector.
Nevertheless, for multiple scattering it presents some inconveniences.
From one scattering event to another, the electric field needs to be written in local basis.
The $\kappa + 1$ scattering event is written in the preceding
$(\hat{\theta}_{\kappa},\hat{\phi}_{\kappa},\hat{k}_{\kappa})$ basis, which depends on the photon history as illustrated by Fig.~\ref{fig:1}.
For instance, to write the scattered field along a scattering sequence in a Monte Carlo scheme, the azimuthal and scattering angles and all the distances along the sequence need to be recorded to perform the right basis transformation back to the laboratory (fixed) frame.
This bookkeeping is necessary when the reversed scattering sequences must be obtained.
This representation becomes specially cumbersome for the calculation of the coherent backscattering enhancement (weak localization of light)~\cite{martinez:5:1994,pinheiro_prl1_2000,pinheiro_prl2_2000,pinheiro_j3m_2001,pinheiro_bjp_2001}.

\begin{figure}
\begin{center}
\includegraphics[width=\columnwidth]{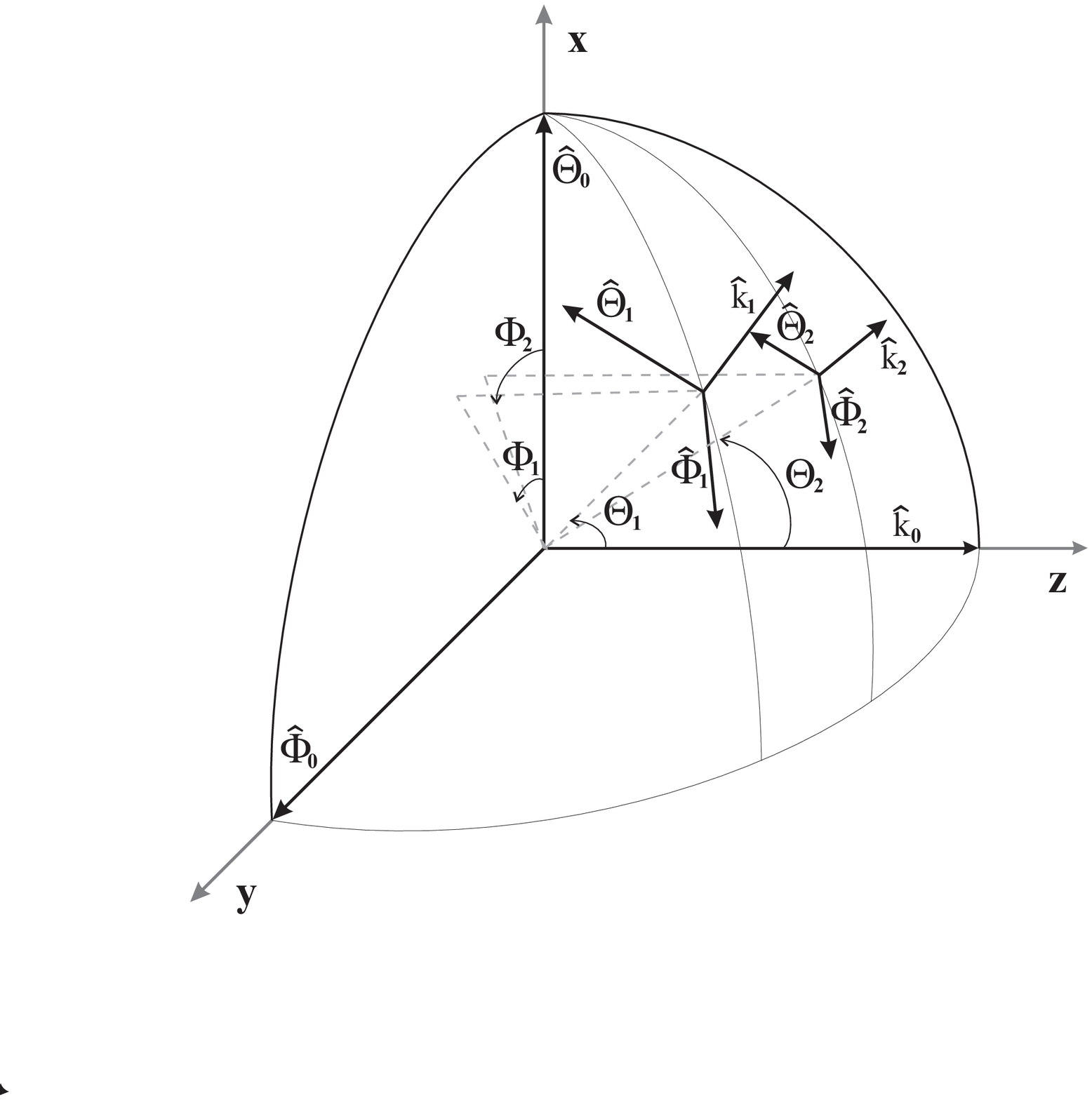}
\end{center}
\vspace{-7cm}
\caption{Angles in the Chandrasekhar-Sekera representation along the scattering sequence of Fig.~\ref{fig:1}.}
\label{fig:2}
\end{figure}

The difficulties presented above can be solved using the Chandrasekhar-Sekera representation~\cite{chandrasekhar_book,sekera}.
Consider for example the $\kappa^{\mbox{th}}$ scattering event in a given scattering sequence.
As previously, an incident monochromatic (plane) wave with wavevector $\vec{k}_{\kappa}$ is considered and one is interested in the scattered field along the direction $\vec{k}_{\kappa+1}$.
Note that in this case the incident direction is arbitrary, not necessarily along the $\hat{z}$-direction, as in the scattering plane representation.
In a distance $R_{\kappa + 1}$ from the center of the sphere, with $(ka)^2/kR_{\kappa + 1} \ll 1$ (far-field approximation), the scattered field is nearly transverse and the electric field lies on the plane formed by the spherical basis vectors $\hat{\Theta}_{\kappa + 1}$ and $\hat{\Phi}_{\kappa + 1}$.
The spherical basis $(\hat{\Theta}_{\kappa}, \hat{\Phi}_{\kappa}, \hat{k}_{\kappa})$ is defined by the angle $\Theta_{\kappa}$, which is obtained from the projection of the wavevector $\vec{k}_{\kappa}$ along the $\hat{z}$-direction $(\cos \Theta_{\kappa} = \hat{k}_{\kappa} \cdot \hat{z})$ and by the azimuthal angle $\Phi_{\kappa}$, which is counted from $\hat{x}$ to the projection of $\vec{k}_{\kappa}$ onto $xy-$plane, relative to the  laboratory frame, as shown in Fig.~\ref{fig:2}.

Equivalently to the scattering plane representation, one can write the scattered electric field $\vec{E}_{\kappa + 1}$ along the directions $(\hat{\Theta}_{\kappa + 1}, \hat{\Phi}_{\kappa + 1})$ as a function of the incident field $\vec{E}_{\kappa}$ along the directions $(\hat{\Theta}_{\kappa}, \hat{\Phi}_{\kappa})$:
\label{eq:etr_fld}
\begin{equation}
\label{eq:ef}
\vec{E}_{\kappa + 1}(\hat{k}_{\kappa + 1}) = -\imath \; \frac{e^{- \imath k R_{\kappa + 1}}}{k R_{\kappa + 1}} \;
J(\hat{k}_{\kappa + 1},\hat{k}_{\kappa}) \;
\vec{E}_{\kappa}(\hat{k}_{\kappa}) \; ,
\end{equation}
where the Jones matrix is:
\begin{eqnarray}
\label{eq:jones_csr}
J(\hat{k}_{\kappa + 1},\hat{k}_{\kappa}) & = &
\left[
\begin{array}{cc}
 (l,l) & (l,r) \\
-(l,r) & (l,l)
\end{array}
\right]
\left[
\begin{array}{cc}
X_1 &  0  \\
 0  & X_2
\end{array}
\right] +  \nonumber \\ & &
\left[
\begin{array}{cc}
 -(r,l) & (r,r) \\
  (r,r) & (r,l)
\end{array}
\right]
\left[
\begin{array}{cc}
 0   & X_1  \\
X_2  &  0
\end{array}
\right] \; ,
\\
(l,l) & = & \sqrt{(1 - \mu_{\kappa}^2)(1 - \mu_{\kappa + 1}^2)} + \nonumber \\
      &   & \mu_{\kappa}\mu_{\kappa+1} \cos(\Delta \Phi_{\kappa,\kappa + 1})  \; ,\label{eq:ll} \\
(l,r) & = & - \mu_{\kappa} \sin(\Delta \Phi_{\kappa,\kappa + 1})   \; , \label{eq:lr} \\
(r,l) & = &  \mu_{\kappa+1} \sin(\Delta \Phi_{\kappa,\kappa + 1})  \; , \label{eq:rl} \\
(r,r) & = & \cos(\Delta \Phi_{\kappa,\kappa + 1})  \; , \label{eq:rr} \\
\label{eq:az_angle}
\Delta \Phi_{\kappa,\kappa + 1} & = & \Phi_{\kappa} - \Phi_{\kappa + 1} \; , \\
\label{eq:cosd}
\mu_{\kappa} & = &  \cos \Theta_{\kappa} \; ,
\end{eqnarray}
where the azimuthal scattering angle is $\Delta \Phi_{\kappa,\kappa + 1}$ (equivalent to $\phi$ in the scattering plane representation).
The notation of Eq.~(218) in Ref.~\cite[p. 38-43]{chandrasekhar_book} has been kept in this calculation and the geometrical interpretation obtained from the spherical triangle shown in Fig.~8 of Ref.~\cite[p. 39]{chandrasekhar_book}.
The modified scattering amplitudes, which depend only on the  cosine of the scattering angle $\mu$ due to the spherical symmetry of the scatterer, are defined by
\begin{eqnarray}
X_1(\mu) & = &  \frac{S_{\perp}(\mu) - \mu S_{\parallel}(\mu)}{1 - \mu^2} \; ,
\label{eq:x1_x2:1} \\
X_2(\mu) & = &  \frac{S_{\parallel}(\mu) - \mu S_{\perp}(\mu)}{1 - \mu^2} \; ,
\label{eq:x1_x2:2} \\
\label{eq:sct_angle}
\nonumber
\mu & = & \cos \theta = (l,l)(r,r) - (l,r)(r,l) \\
    & = & \sqrt{(1 - \mu_{\kappa}^2)(1 - \mu_{\kappa + 1}^2)} \cos(\Delta \Phi_{\kappa,\kappa + 1}) +  \mu_{\kappa}\mu_{\kappa+1} \;
    ,
\end{eqnarray}
with the scattering amplitudes given by Eqs.~(\ref{eq:s1s2:1}) and~(\ref{eq:s1s2:2}).
Because of the basis transformations, the Jones matrix in the Chandrasekhar-Sekera representation Eq.~(\ref{eq:jones_csr}) cannot simply be written as product of a scattering  by a rotation matrix as in the scattering plane representation [Eq. (\ref{eq:jones_spr})].
Nevertheless, writing  $\hat{k}_{\kappa} = \hat{z}$, $\hat{\Theta}_{\kappa} = \hat{x}$, and $\hat{\Phi}_{\kappa} = \hat{y}$, then $\Phi_{\kappa} = 0$, and $\Theta_{\kappa} = 0$ for the incident direction and $\Phi_{\kappa + 1} = \phi$, and $\Theta_{\kappa + 1} = \theta$, then $\Delta \Phi_{\kappa,\kappa +_ 1} = - \phi$, $\mu_{\kappa} = 1$, and $\mu_{\kappa + 1} = \cos \theta$, the scattering plane representation can be obtained as a particular case.

Although the Chandrasekhar-Sekera representation is more complex than the scattering plane representation, it is suitable for numerical calculation in the multiple scattering context since it already includes the basis transformation in itself.
In this representation, given $\hat{k}_{\kappa}$ and $\hat{k}_{\kappa + 1}$ in the laboratory frame, the electric field along the scattering sequence can be promptly obtained, even in the case where the reverse sequences must be calculated such as in the weak localization of eletromagnetic waves.

\section{Non-commutability of Limiting Cases}
\label{sec:limit_problems}

To calculate the scattering of light by polydispersions, particle size may range from the small particle to large sphere limit.
Numerical problems with Eqs. (\ref{eq:x1_x2:1}) and (\ref{eq:x1_x2:2}) may occur since forward/backward scattering events $(\mu \rightarrow \pm 1)$ have non-vanishing probability.

Backward/forward scattering are important in practice.
For instance, the phase functions are very forwardly peaked ($\mu \rightarrow  1$) in the Mie regime.
In lidar calculation, the interest is in the backward scattering events ($\mu \rightarrow -1$, in general these scattering events are forced to occur by Metropolis variance-reduction methods in a Monte Carlo scheme).

Since these difficulties have never been reported in the literature, they are explicitly developed here, where we consider the particles to be magnetic.
The non-magnetic approach, which is widely applied in the optical frequency ranges for all materials, can be readily obtained by assuming $\widetilde{\mu} = \mu_s/\mu_m = 1$ in our expressions.

A simple way to obtain the limiting values, which works properly for the optical theorem, is to rewrite Eqs. (\ref{eq:x1_x2:1}) and (\ref{eq:x1_x2:2}) as: $X_1(\mu)  =   [S_{\perp}(\mu)     - \mu S_{\parallel}(\mu)]/[(1 + \mu)(1 - \mu)]$ and $X_2(\mu)  =   [S_{\parallel}(\mu) - \mu S_{\perp}    (\mu)]/[(1 + \mu)(1 - \mu)]$.

In the forward/backward limiting cases of $\mu \rightarrow \pm 1$,  $S_{\perp}(1) = S_{\parallel}(1) = S(1)$, $S_{\perp}(-1)  =  -S(-1)$ and   $S_{\parallel}(-1) = S(-1)$, with $S(\pm 1)$ given by Eq. (\ref{eq:s0}), and one can write: $X_1(\mu \rightarrow \pm 1)  =  \pm S(\pm 1)/2$ and  $X_2(\mu \rightarrow \pm 1)  =  S(\pm 1)/2$.
In short, $X_1(\pm 1) = \pm X_2(\pm 1)$.

Now let us consider the limit $ka \rightarrow 0$ (small-particle limit), and the leading Mie coefficients are $a_1 = 2\imath \alpha (ka)^3/3$ and $b_1=2\imath \beta (ka)^3/3$, with $\alpha = (\widetilde{\epsilon} - 1)/(\widetilde{\epsilon} + 2)$ and $\beta= (\widetilde{\mu} - 1)/(\widetilde{\mu} + 2)$~\cite{kerker:1983}.
From Eqs.~(\ref{eq:s1s2:1}) and (\ref{eq:s1s2:2}), the scattering amplitudes are $S_{\perp}(\mu) = \imath (\alpha+\mu\beta) (ka)^3$ (independent of $\mu$ when $\widetilde{\mu}=1$) and $S_{\parallel}(\mu) = \imath (\mu\alpha+\beta) (ka)^3$, and so that: $X_1(\mu)  =   \imath \alpha (ka)^3$ and $X_2(\mu)  = \imath \beta (ka)^3$.
For $\beta=0$ and $\mu\to1$, this implies $X_1(1)  =  S_{\perp} = S_{\parallel}$ and $X_2(1)  =  0$, leading to $X_1(\pm 1) \neq \pm X_2(\pm 1)$.

Formally, in the non-magnetic approach ($\beta=0$), $X_1$ tends to zero, as $ka$ tends to zero, while $X_2$ vanishes.
Taking the limit $\mu \rightarrow \pm 1$ and then $ka \rightarrow 0$ the result is different than taking $ka \rightarrow 0$ and then $\mu \rightarrow \pm 1$.
These limits do not commute, which is a non-physical result.
Furthermore, the order the limits are taken affects numerical results.

Once this problem has been noticed, the solution is obvious:  expand the scattering amplitudes to higher-order terms as function of  $\mu=\cos \theta$.
Because of the presence of a $1 - \mu^2 = \sin^2 \theta$ in the denominator of $X_1(\mu)$ and $X_2(\mu)$ [Eqs. (\ref{eq:x1_x2:1}) and (\ref{eq:x1_x2:2})], the consideration up to fourth-order terms in $\theta$ is mandatory here, in contrast to the use of the optical theorem.

For $\mu \rightarrow \pm 1$, using Eqs.~(\ref{eq:scat_amp:1}), (\ref{eq:scat_amp:2}), (\ref{eq:x1_x2:1}) and (\ref{eq:x1_x2:2}),  one obtains:
\begin{eqnarray*}
X_1(\mu \rightarrow \pm 1) & = & \pm \frac{S(\pm 1)}{2} \pm \Delta^{(\pm)}\ , \\
X_2(\mu \rightarrow \pm 1) & = & \frac{S(\pm 1)}{2} - \Delta^{(\pm)} \; ,
\end{eqnarray*}
where $\Delta^{(\pm)}$ is given by Eq.~(\ref{eq:delta}) and which corrects the problem of non-commutability of limits.
For small particle scattering, $\Delta^{(\pm)} = \imath (\mp\alpha-\beta) (ka)^3/2$, which leads to the correct values: $X_1 = \imath \alpha (ka)^3$ and $X_2 = \imath \beta (ka)^3$.


\section{The New Set of Basis Functions}
\label{sec:the_new_set_of_basis_functions}

An effective manner to solve the non-commutability of the limiting cases is to use the definitions of the scattering amplitudes [Eqs. (\ref{eq:s1s2:1}) and (\ref{eq:s1s2:2})] and rewrite $X_1(\mu)$ and $X_2(\mu)$ in Eqs. (\ref{eq:x1_x2:1}) and (\ref{eq:x1_x2:2}) as:
\begin{eqnarray}
X_1(\mu)  & = & \sum_{n = 1}^{\infty} \frac{2n + 1}{n(n+1)} \; [a_n \chi_{n}^{(1)}(\mu) + b_n \chi_{n}^{(2)}(\mu) ] \; ,
\label{eq:nf1} \\
X_2(\mu)  & = & \sum_{n = 1}^{\infty} \frac{2n + 1}{n(n+1)} \; [a_n \chi_{n}^{(2)}(\mu) + b_n \chi_{n}^{(1)}(\mu) ] \; ,
\label{eq:nf2} \\
\chi_{n}^{(1)}(\mu) & = & \frac{\pi_n(\mu) - \mu \tau_n(\mu)}{1 - \mu^2} = \pi_n(\mu) + \mu \pi^{\prime}_n(\mu) \; ,
\label{eq:chi:1} \\
\chi_{n}^{(2)}(\mu) & = & \frac{\tau_n(\mu) - \mu \pi_n(\mu)}{1 - \mu^2} = - \pi^{\prime}_n(\mu) \; ,
\label{eq:chi:2}
\end{eqnarray}
where Eqs. (\ref{eq:pi}) and (\ref{eq:tau}) have been used.

Instead of calculating the scattering amplitudes $S_{\perp}(\mu)$ and $S_{\parallel}(\mu)$, $X_1(\mu)$ and $X_2(\mu)$ can be directly calculated from the basis functions $\chi_{n}^{(1)}(\mu)$ and $\chi_{n}^{(2)}(\mu)$.
The consideration of Eqs.~(\ref{eq:chi:1}) and~(\ref{eq:chi:2}) implies to an important achievement, since the term $\sin^2 \theta$ that appears in the denominator of Eqs.~(\ref{eq:x1_x2:1}) and~(\ref{eq:x1_x2:2}) cancels out when one writes $\tau_{n}(\mu)  =  \mu \pi_n(\mu) - (1 - \mu^2) \pi^{\prime}_n(\mu)$ [Eq. (\ref{eq:tau})].
In this way, no asymptotic behavior for $X_{1}(\mu)$ and $X_{2}(\mu)$ must be explicitly calculated numerically, avoiding numerical tests, which increase computer time.
This procedure achieves directly the limiting cases, independently of the order these limits are taken.

The functions $\pi_n(\mu)$ and $\pi^{\prime}_n(\mu)$ can be obtained numerically by stable recursion relations \cite{kerker_book,wiscombe:1980}
\begin{eqnarray}
t_{n - 1}(\mu)        & = & (2n - 1) \pi_{n-1}(\mu)
\label{eq:leg_pol_1} \\
\pi^{\prime}_{n}(\mu) & = & t_{n-1}(\mu) + \pi^{\prime}_{n-2}(\mu)
\label{eq:leg_pol_2} \\
\pi_{n}(\mu)          & = & \frac{\mu t_{n-1}(\mu) - n \pi_{n-2}(\mu)}{n - 1} \; ,
\label{eq:leg_pol_3}
\end{eqnarray}
which are initialized with $\pi_{0}(\mu) = 0$, $\pi_{1}(\mu) = 1$, $\pi_{2}(\mu) = 3 \mu$,
$\pi_{1}^{\prime}(\mu) = 0$ and $\pi_{2}^{\prime}(\mu) = 3$.

A further improvement can be achieved in numerical calculations.
Instead of using $X_1(\mu)$ and $X_2(\mu)$, one writes:
\begin{eqnarray}
\nonumber 
X_{\pm}(\mu) & = & \frac{X_{1}(\mu) \pm X_{2}(\mu)}{2} \\
             & = & \frac{1}{2} \sum_{n = 1}^{\infty} \frac{2n + 1}{n(n+1)} \;
                 (a_n \pm b_n)\chi_n^{(\pm)}(\mu)
\label{eq:xpm} \\
\chi_n^{(\pm)}(\mu) & = & \chi_n^{(1)}(\mu) \pm \chi_n^{(2)}(\mu) \; ,
\label{eq:chi_pm_2}
\end{eqnarray}
which is an adaptation of the Wiscombe's method \cite{wiscombe:1980}.
For small particle scattering $X_{\pm}(\mu)  =  \imath (\alpha\pm\beta) (ka)^3/2$, which leads to the correct limits  $X_1 =  \imath \alpha (ka)^3$ and $X_2 = \imath\beta(ka)^3$.

In brief the new algorithm works as follows.
Given the size parameter $ka$, the relative refraction index $m$ and $\widetilde{m}$ (for magnetic scattering), one calculates the Mie coefficients $a_n$ and $b_n$ and store the terms $(2n + 1)(a_n \pm b_n)/[2n(n+1)]$ [Eq. (\ref{eq:xpm})] in arrays \cite{wiscombe:1980}.
Given the input $\hat{k}_{\kappa}$ and output $\hat{k}_{\kappa + 1}$ directions, proceed as follows:
\begin{enumerate}
\item obtain the scattering angle as given by Eq. (\ref{eq:sct_angle});
\item calculate the ``Legendre'' polynomials by the recursion relationships of
Eqs.~(\ref{eq:leg_pol_1},)~(\ref{eq:leg_pol_2}) and~(\ref{eq:leg_pol_3},);
\item the basis functions $\chi_n^{\pm}(\mu)$ are obtained by  Eq.~(\ref{eq:chi_pm_2});
\item $X_{\pm}(\mu)$, $X_1(\mu)$ and $X_{2}(\mu)$ are given by Eqs.~(\ref{eq:xpm}),~(\ref{eq:nf1})  and~(\ref{eq:nf2});
\item the azimuthal angle is calculated with Eq. (\ref{eq:az_angle});
\item the director cosines are obtained by Eq. (\ref{eq:cosd});
\item use Eqs. (\ref{eq:ll}), (\ref{eq:lr}), (\ref{eq:rl}) and (\ref{eq:rr})
to obtain $(l,l)$, $(l,r)$, $(r,l)$ and $(r,r)$, respectively;
\item the results of the above items permit the calculations of the Jones matrix
with Eq. (\ref{eq:jones_csr}),
\item and finally, at the distance $R_{\kappa + 1}$ from the center of the sphere
along $\hat{k}_{\kappa + 1}$, the scattered electric field is given by Eq. (\ref{eq:ef}).
\end{enumerate}

The calculations are performed on the field level.
To obtain the Stokes intensities ($I$, $Q$, $U$ and $V$) use the coherence matrix $\vec{E}_{\kappa + 1}^{\dag} \vec{E}_{\kappa + 1}$~\cite{martinez:5:1994}.

For a fixed particle size (monodispersions), the functions $\chi_n^{(\pm)}(\mu)$ are calculated each time the angles of source and detection are modified, or when a new scattering event occurs in a Monte Carlo sequence.
In this case, it is suitable to tabulate $X_{\pm}(\mu)$ [Eq.~(\ref{eq:xpm})],
in uneven partitions.
Higher resolution are required for $|\mu|$ close to unity~\cite{nussenzveig_book},  roughly $\theta < ka$ in the forward case  and $\pi - \theta <  (ka)^{4/3}$ in the backward case.
To consider polydispersions,  the size-distribution function must be taken into account.
Some of these size-distribution functions are found in Ref.~\cite{deirmendjian_book} and an implementation for large size parameters is found in Ref.~\cite{wolf_2004}.
In this case, values of $\langle |X_{+}(\mu)|^2 \rangle$,
$\langle |X_{-}(\mu)|^2 \rangle$, $\mbox{Re}[\langle X_{+}(\mu)X_{-}^{*}(\mu) \rangle]$ and
$\mbox{Im}[\langle X_{+}(\mu)X_{-}^{*}(\mu) \rangle]$ should be tabulated, where $\langle \ldots \rangle$ refers to the size-distribution average.

\section{Conclusion}
\label{sec:conclusion}

The Chandrasekhar-Sekera representation is more appropriate than the scattering plane representation to be considered when multiple Mie scattering of light is taken into account.
Nevertheless, care must be taken when this representation is used because of the non-commutability of the limiting cases of Rayleigh and forward/backward scattering.
It must be corrected considering the higher-order terms in the scattering angle in the scattering amplitude expansions.
A new set of basis functions for the calculation of the angular functions in the Mie scattering in the Chandrasekhar-Sekera representation has been presented.
An effective algorithm has been implemented using new basis functions, which have been introduced.

Furthermore, the use of a generalization (to associated Legendre polynomials) of $\chi_{n}^{(1)}$ and $\chi_{n}^{(2)}$ can be used for non-spherical scattering, such as in the extended boundary condition presented in Ref.~\cite{barber_book}.
This generalization, among the generalization for the spherical Henkel functions, allows the writing of the scattering amplitudes of non-spherical particles in terms of matrices, which are suitable for numerical calculations.
These calculations will be reported in future.

\section*{Acknowledgement}

ASM acknowledges the support from the Brazilian agency CNPq (303990/2007-4).
TJA acknowledges the support from the Brazilian agency FAPESP (2010/10052-0).
The authors thank the very stimulating discussions with Felipe Arruda Pinheiro. 



\end{document}